\documentclass[aps,pra,reprint,amsmath,amssymb,nofootinbib,floatfix,showkeys]{revtex4-2}

\usepackage[T1]{fontenc}
\usepackage{lmodern}
\usepackage{graphicx}
\usepackage{bm}
\usepackage{mathtools}
\usepackage{microtype}
\usepackage[hidelinks]{hyperref}
\hypersetup{pdftitle={Four-Sector Propagation-Trapping Decomposition and Inverse Design of Initial States in Grover-Coined Quantum Walks on Multiedge Path Graphs},pdfauthor={Hamza Bougroura}}

\newcommand{\ones}{\mathbf{1}}
\newcommand{\zeros}{\mathbf{0}}

\newcommand{\rv}{\mathbf r}
\newcommand{\vt}{V_{\mathrm{tot}}}
\newcommand{\pair}[2]{\begin{pmatrix}#1\\#2\end{pmatrix}}

\newcommand{\stackblocks}[2]{\left(\begin{gathered}#1\\[4pt]#2\end{gathered}\right)}

\newcommand{\tabblocks}[2]{\bigl((#1),(#2)\bigr)^T}
\begin{document}

% Compact but readable display spacing, avoiding the oversized gaps seen in the Word export.
\setlength{\abovedisplayskip}{5.5pt plus 1.5pt minus 1.5pt}
\setlength{\belowdisplayskip}{5.5pt plus 1.5pt minus 1.5pt}
\setlength{\abovedisplayshortskip}{3.5pt plus 1pt minus 1pt}
\setlength{\belowdisplayshortskip}{3.5pt plus 1pt minus 1pt}
\setlength{\textfloatsep}{9pt plus 2pt minus 2pt}
\setlength{\floatsep}{8pt plus 2pt minus 2pt}
\setlength{\intextsep}{8pt plus 2pt minus 2pt}

\title{Four-Sector Propagation--Trapping Decomposition and Inverse Design of Initial States in Grover-Coined Quantum Walks on Multiedge Path Graphs}
\author{Hamza Bougroura}
\email{hamza.bougroura@univ-batna.dz}
\affiliation{Department of Fundamental Education in Matter Sciences, Faculty of Matter Sciences, University of Batna 1, Batna 05000, Algeria}
\affiliation{Radiation Physics and Interaction with Matter Laboratory (PRIMALAB), Faculty of Matter Sciences, University of Batna 1, Batna 05000, Algeria}
\date{November 7, 2026}

\begin{abstract}
Controlling propagation and trapping through initial-state preparation requires an explicit relation between the initial amplitudes and the corresponding directional probabilities. We establish this relation for a discrete-time Grover-coined quantum walk on a multiedge path graph with even degree $d=2M\ge4$ and two equal directional blocks. Each block of an arbitrary local state is expressed as its mean component plus a zero-sum deviation. The Grover coin performs a pure exchange of the two block means and reverses the sign of each zero-sum deviation. Together with the graph routing, this gives a unique orthogonal decomposition into four sectors: propagation in opposite directions and bounded trapping on either side of the launch vertex. We derive the four sector probabilities and solve the inverse problem, constructing initial states for any prescribed nonnegative probability quadruple that sums to unity. Explicit representatives are given for $d=4$ and $d=6$. Sector-resolved simulations reproduce the predicted motion, conservation of sector weights, and local probability additivity, including repeated spatial overlaps. We also derive the local Grover action for an arbitrary partition into blocks of possibly unequal sizes, explicitly describing how their means are redistributed while each zero-sum deviation remains within its block and changes sign. The framework provides direct control of propagation and trapping within one fixed graph and walk operator through the preparation of the initial state.
\end{abstract}


\maketitle
\raggedbottom

\section{Introduction}

Quantum walks provide coherent counterparts of classical random walks, with transport governed by amplitudes and interference~\cite{Aharonov1993}. Early discrete-time formulations established fast spreading on lines and graphs, extensions to higher dimensions, and dependence on the coin and initial state~\cite{Ambainis2001,Aharonov2001,Mackay2002,Konno2002,Tregenna2003}; broader developments and applications are reviewed in Ref.~\cite{Venegas2012}. In a coined discrete-time quantum walk (DTQW), the graph and shift specify the available routes, the coin mixes amplitudes locally, and the initial state selects the populated dynamical components. Their compatibility determines propagation, recurrence, trapping, and state transfer.

Symmetry strongly affects transport. Graph symmetries can produce invariant subspaces that prevent particular initial states from reaching a target, and quotient graphs expose the resulting reduced dynamics~\cite{Krovi2006,Krovi2007}. Initial and target states respecting the graph symmetry can improve transport efficiency~\cite{Bougroura2016}. Structural descriptions connect discrete-time walks with graph embeddings and symmetric factorizations~\cite{Godsil2019}, while coin symmetry can modify symmetry-induced hitting-time behavior~\cite{Prabhu2022}. These results show why transport must be examined through the combined graph, coin, and state preparation.

Perfect state transfer (PST) provides a precise transport benchmark. Early spin-network proposals established high-fidelity communication and exact transfer in suitable chains~\cite{Bose2003,Christandl2004}, followed by broader analyses relating transfer to graph structure and periodicity~\cite{Kendon2011,Godsil2012}. In discrete time, deterministic Grover-coined wires and position-dependent coins provide transfer constructions~\cite{Lovett2010,Kurzynski2011}. Transfer has also been investigated on modified cycles and with coin-dependent recovery operations~\cite{Barr2014,Yalcinkaya2015}, through programmed coin operations~\cite{Zhan2014}, and through dynamical graph control and initial-state preparation~\cite{Nitsche2016}. Search-based constructions address highly symmetric and complete bipartite graphs~\cite{Stefanak2016,Stefanak2017}. The complete-graph Szegedy construction in Ref.~\cite{Stefanak2016} approaches unit fidelity as the graph size increases; this differs from exact finite-size PST. Spectral studies identify circulant families with PST and necessary conditions for vertex-associated transfer in Grover walks~\cite{Zhan2019,Kubota2022}. Pretty-good state transfer concerns arbitrarily close transfer on a fixed graph~\cite{ChanZhan2023}. Recent classifications further constrain PST in low-valency circulant and Cayley Grover walks~\cite{KubotaYoshino2025,Bhakta2026}.

Localization and trapping provide a complementary perspective. The two-dimensional Grover walk exhibits localization associated with spectral degeneracy and initial-state dependence~\cite{Inui2004}. A one-dimensional three-state walk supports persistent probability near the origin~\cite{Inui2005}, while a phase defect can trap a walker on a line~\cite{Wojcik2012}. Ambainis et al.~\cite{Ambainis2016Oscillatory} characterized flip states by vanishing incoming and outgoing amplitude sums, established their sign reversal under the Grover coin and exact two-step recurrence under flip-flop routing, and connected oscillatory localization to electrical-network properties. We independently derived the corresponding local trapping relations before becoming aware of that work. Here, we use a local mean–deviation formulation to unify propagation and trapping in four orthogonal sectors.
We study a multiedge path graph with local degree
\begin{equation}
d=2M,\qquad M\ge2,
\label{eq:d2M}
\end{equation}
and two equal directional blocks of $M$ channels. Equal-amplitude transfer between equally sized input and output blocks underlies deterministic Grover-coined wires~\cite{Lovett2010}. Here the geometric block size matches the factor $2/d=1/M$ in the Grover coin, turning each block sum into its mean. We use this compatibility to characterize both propagation and trapping for an arbitrary local initial state. Each block separates into a uniform mean component and a zero-sum deviation, producing four mutually orthogonal sectors. Their probabilities give the complete local transport content relative to the chosen partition. Inverting these relations yields a family of initial states for any admissible set of four probabilities, without changing the graph or walk operator. The numerical study checks the predicted spatial motion and local probability additivity when sector contributions meet repeatedly.

Section~\ref{sec:model} introduces the model and its stepwise dynamics. Section~\ref{sec:characterization} derives the elementary modes and four-sector decomposition. Section~\ref{sec:probabilities} develops the probabilities and inverse construction, and Sec.~\ref{sec:examples} gives explicit examples. Section~\ref{sec:numerical} presents the numerical verification, followed by the conclusion. Appendix~\ref{app:blocks} derives the local Grover action for arbitrary block partitions.

\section{Model and stepwise dynamical picture}
\label{sec:model}

We consider a multiedge path whose bulk vertices have even degree $d=2M$, with $M\ge2$. The graph geometry partitions the local channels into $B_1=\{1,\ldots,M\}$ and $B_2=\{M+1,\ldots,2M\}$, associated with the left and right neighboring vertices, respectively. The local Grover coin is
\begin{equation}
 G_d=\frac{2}{d}J_d-I_d=\frac1M J_d-I_d,
 \label{eq:Grover}
\end{equation}
where $J_d$ is the $d\times d$ all-ones matrix. For a finite realization with $N$ vertices, one step consists of a coin operation followed by the shift,
\begin{equation}
 U=S\bigl(\mathbb I_N\otimes G_d\bigr).
 \label{eq:step}
\end{equation}
For $i=1,\ldots,M$, the bulk channel routing is
\begin{equation}
\begin{aligned}
 S|n,i\rangle&=|n-1,i+M\rangle,\\
 S|n,i+M\rangle&=|n+1,i\rangle.
\end{aligned}
\label{eq:shiftrouting}
\end{equation}
These formulas describe evolution before boundary encounters; the periodic closure used in the numerical study is specified in Sec.~\ref{sec:numerical}.

Let the walker initially be localized at vertex $0$, with local coin vector
\begin{equation}
 V=\bigl((z_1,\ldots,z_M),(z_{M+1},\ldots,z_{2M})\bigr)^T.
 \label{eq:localV}
\end{equation}
The inner parentheses distinguish the two blocks, with indices $1,\ldots,M$ and $M+1,\ldots,2M$. The case \mbox{$d=4$ ($M=2$)} gives a direct illustration of the local mechanism. Define the two block means by
\begin{equation}
 \bar{x}=\frac{z_1+z_2}{M},\qquad
 \bar{y}=\frac{z_3+z_4}{M},\qquad M=2.
 \label{eq:means4}
\end{equation}

Adding and subtracting the corresponding block mean in each component gives
\begin{widetext}
\begin{equation}
 V=\stackblocks{\pair{z_1}{z_2}}{\pair{z_3}{z_4}}
 =\stackblocks{\pair{\bar{x}+z_1-\bar{x}}{\bar{x}+z_2-\bar{x}}}{\pair{\bar{y}+z_3-\bar{y}}{\bar{y}+z_4-\bar{y}}}
 =\stackblocks{\pair{\bar{x}+(z_1-\bar{x})}{\bar{x}+(z_2-\bar{x})}}{\pair{\bar{y}+(z_3-\bar{y})}{\bar{y}+(z_4-\bar{y})}}.
 \label{eq:meanplus4}
\end{equation}
The inner column-vector parentheses identify the two blocks; the parentheses within each entry distinguish its deviation from the common mean. The deviations sum to zero separately within each block. Since the total amplitude sum is $M(\bar{x}+\bar{y})$, with $M=2$, the Grover transformation becomes
\begin{equation}
 G_4V=G_4\stackblocks{\pair{\bar{x}+(z_1-\bar{x})}{\bar{x}+(z_2-\bar{x})}}{\pair{\bar{y}+(z_3-\bar{y})}{\bar{y}+(z_4-\bar{y})}}
 =\stackblocks{\pair{\bar{y}-(z_1-\bar{x})}{\bar{y}-(z_2-\bar{x})}}{\pair{\bar{x}-(z_3-\bar{y})}{\bar{x}-(z_4-\bar{y})}}.
 \label{eq:Grover4means}
\end{equation}
\end{widetext}

\begin{figure*}[!t]
 \centering
 \includegraphics[width=\textwidth]{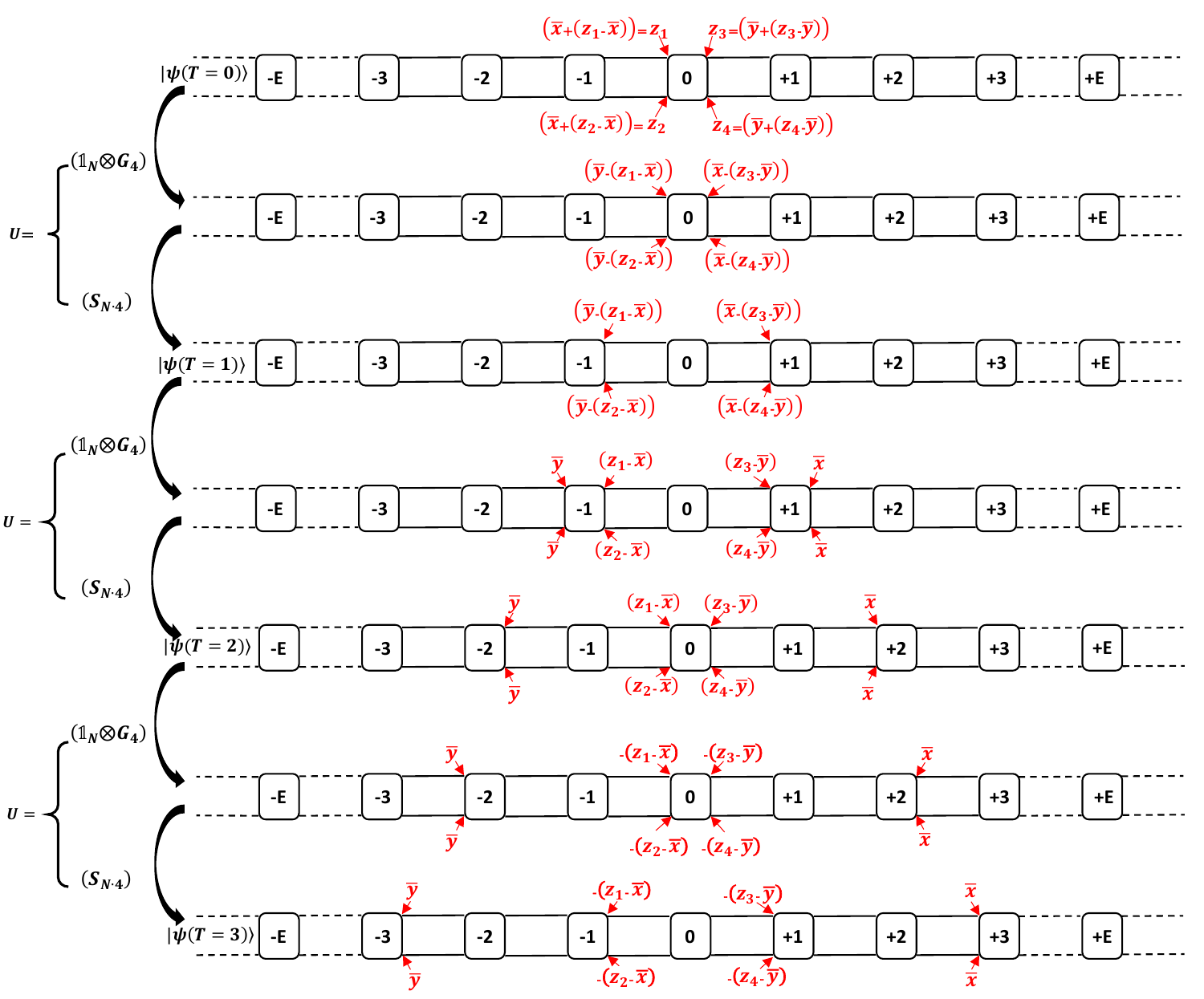}
 \caption{Stepwise coin--shift evolution of an arbitrary local state on the $d=4$ multiedge path. The illustrated segment contains $N=2E+1$ vertices, with the initial state localized at vertex $0$. Here $M=d/2=2$, and the block means are $\bar{x}=(z_1+z_2)/M$ and $\bar{y}=(z_3+z_4)/M$. In the initial row, each amplitude is written as its block mean plus its deviation: $z_i=\bar{x}+(z_i-\bar{x})$ for $i=1,2$, and $z_i=\bar{y}+(z_i-\bar{y})$ for $i=3,4$. Each time step consists of $\mathbb I_N\otimes G_4$ followed by the shift, with intermediate coin stages displayed explicitly. The coin exchanges the block means and reverses the deviations. Under the depicted routing, $\bar{x}$ propagates rightward and $\bar{y}$ leftward, while the first- and second-block deviations alternate between $0$ and $-1$, and between $0$ and $+1$, respectively. The first three steps are shown before boundary encounters.}
 \label{fig:evolution}
\end{figure*}

Thus, the coin exchanges the two block means and reverses the sign of each deviation. Figure~\ref{fig:evolution} follows the successive coin and shift operations through the first three steps. Under the depicted routing, the mean contributions move away from the origin, whereas the deviations alternate between the origin and one adjacent vertex. The next section derives the elementary conditions underlying these behaviors and extends this representation to arbitrary even degree.

\section{Characterization and decomposition of fundamental transport modes}
\label{sec:characterization}

\subsection{Elementary perfect-transfer and perfect-trapping modes}

Consider a nonzero elementary input supported within one selected channel set $B$, where $B=B_1$ or $B=B_2$. The complementary block is denoted by $B^c=\{1,\ldots,2M\}\setminus B$: thus $B_1^c=B_2$ and $B_2^c=B_1$. The superscript $c$ denotes set complement. The input satisfies $z_j=0$ for every $j\in B^c$. Let
\begin{equation}
 s=\sum_{i=1}^{2M}z_i=\sum_{i\in B}z_i.
 \label{eq:generalVsum}
\end{equation}
The componentwise Grover transformation is
\begin{equation}
 (G_dV)_i=\frac{s}{M}-z_i.
 \label{eq:componentGrover}
\end{equation}
The following conditions characterize complete transfer to the complementary block and complete suppression of that transfer.

\subsubsection{Elementary perfect-transfer modes}

Let $K=\{i\in B:z_i\ne0\}\subseteq B$ be the set of indices of the initially nonzero input amplitudes, and let $k=|K|$ denote the number of these indices. Perfect transfer requires the corresponding amplitudes to vanish after the coin operation:
\begin{equation}
 \frac{s}{M}-z_i=0
 \quad\Longrightarrow\quad z_i=\frac{s}{M},\qquad \forall i\in K.
 \label{eq:transferrequirements}
\end{equation}
The initially nonzero amplitudes are therefore all equal. Denote their common value by $A$. Since $s=kA$, a nonzero input requires
\begin{equation}
 A=\frac{kA}{M},\quad A\ne0
 \quad\Longrightarrow\quad k=M=\frac d2.
 \label{eq:nM}
\end{equation}
Thus, every channel in the selected block is occupied with the same amplitude. The initially empty complementary block acquires $(G_dV)_j=s/M=A$ for all $j\in B^c$. The common amplitude is precisely the mean amplitude of the occupied block:
\begin{equation}
 A=\frac{s}{M}=\frac1M\sum_{i\in B}z_i
 =\begin{cases}
 \bar{x},&B=B_1,\\
 \bar{y},&B=B_2.
 \end{cases}
 \label{eq:commonmean}
\end{equation}
Let $\ones_M=(1,\ldots,1)^T$ and $\zeros_M=(0,\ldots,0)^T$ denote the $M$-dimensional column vectors of ones and zeros. With the routing of Fig.~\ref{fig:evolution}, the elementary propagation states are
\begin{equation}
 V_{P_R}=\bar{x}\begin{pmatrix}\ones_M\\\zeros_M\end{pmatrix},\qquad
 V_{P_L}=\bar{y}\begin{pmatrix}\zeros_M\\\ones_M\end{pmatrix}.
 \label{eq:propstates}
\end{equation}
Their Grover outputs are
\begin{equation}
 G_dV_{P_R}=\bar{x}\begin{pmatrix}\zeros_M\\\ones_M\end{pmatrix},\qquad
 G_dV_{P_L}=\bar{y}\begin{pmatrix}\ones_M\\\zeros_M\end{pmatrix}.
 \label{eq:propswap}
\end{equation}
These expressions also establish sufficiency: a uniform input on either directional block is transferred completely to the other block.

\subsubsection{Elementary perfect-trapping modes}

We retain the same elementary setting: the input is supported within one selected $M$-channel block. Here $M$ is fixed by the graph geometry; individual entries in the selected block may vanish. Suppressing transfer to the complementary block requires
\begin{equation}
 (G_dV)_j=\frac{s}{M}-0=0,\qquad j\in B^c,
 \label{eq:trapempty}
\end{equation}
and therefore $s=\sum_{i\in B}z_i=0$. For every component in the selected block, the coin then gives
\begin{equation}
 (G_dV)_i=\frac{s}{M}-z_i=-z_i,\qquad i\in B.
 \label{eq:trapoccupied}
\end{equation}
Thus, a zero sum within the selected block is necessary and sufficient for the coin to preserve its support and reverse its amplitude pattern. Under the routing of Fig.~\ref{fig:evolution}, the subsequent coin--shift sequence confines this component to the origin and the corresponding neighboring vertex.

Let $w_L,w_R\in\mathbb C^M$ denote zero-sum amplitude patterns within the first and second blocks. The elementary trapping states are
\begin{equation}
\begin{aligned}
 V_{T_L}&=\begin{pmatrix}w_L\\\zeros_M\end{pmatrix},&\ones_M^\dagger w_L&=0,\\[3pt]
 V_{T_R}&=\begin{pmatrix}\zeros_M\\w_R\end{pmatrix},&\ones_M^\dagger w_R&=0.
\end{aligned}
\label{eq:trapstates}
\end{equation}

The zero-sum condition applies separately to the selected block. Preserving the occupied amplitudes without a sign change would instead require
\begin{equation}
 \frac{s}{M}-z_i=z_i\quad(i\in B),\qquad
 \frac{s}{M}=0\quad(j\in B^c).
 \label{eq:plushalfsupport}
\end{equation}
The second condition gives $s=0$, and the first then forces every $z_i$ to vanish. Hence this alternative yields no nonzero elementary trapping state.

\subsection{Decomposition of an arbitrary state vector}

The elementary results identify two structures within each block: a uniform amplitude associated with perfect transfer and a zero-sum amplitude pattern associated with trapping. We now express an arbitrary local vector in terms of these structures.\footnote{An earlier use of amplitude averages and deviations in Grover search with arbitrary initial amplitudes can be found in Biham et al.~\cite{Biham1999}. Here we apply this representation to the directional blocks of a multiedge-path quantum walk to characterize propagation and trapping.}

Write the two unrestricted block vectors as
\[
X=(z_1,\ldots,z_M)^T,\qquad
Y=(z_{M+1},\ldots,z_{2M})^T.
\]
For the first block, write $X=c\ones_M+(X-c\ones_M)$. Requiring the residual to have zero sum determines $c$ uniquely:
\begin{equation}
 \sum_{i=1}^{M}(z_i-c)=0
 \quad\Longleftrightarrow\quad
 c=\frac1M\sum_{i=1}^{M}z_i=\bar{x}.
 \label{eq:firstmeanunique}
\end{equation}
Applying the same argument to the second block gives
\begin{equation}
 \bar{y}=\frac1M\sum_{i=M+1}^{2M}z_i.
 \label{eq:secondmeanunique}
\end{equation}
Thus, the uniform amplitude in each block is its mean. This is the familiar geometric decomposition of points into their centroid and their displacements from it; complex amplitudes can be viewed as points in the complex plane.

Define
\begin{equation}
\begin{aligned}
 \Delta\bar X_M&=X-\bar{x}\ones_M=(z_1-\bar{x},\ldots,z_M-\bar{x})^T,\\[3pt]
 \Delta\bar Y_M&=Y-\bar{y}\ones_M=(z_{M+1}-\bar{y},\ldots,z_{2M}-\bar{y})^T.
\end{aligned}
 \label{eq:deltas}
\end{equation}
By construction, $\ones_M^\dagger\Delta\bar X_M=\ones_M^\dagger\Delta\bar Y_M=0$. Every local vector therefore has the representation
\begin{equation}
 V=\begin{pmatrix}X\\Y\end{pmatrix}
 =\begin{pmatrix}
 \bar{x}\ones_M+\Delta\bar X_M\\
 \bar{y}\ones_M+\Delta\bar Y_M
 \end{pmatrix}.
 \label{eq:Vdecomposition}
\end{equation}
Because $s=M(\bar{x}+\bar{y})$, its Grover transformation is
\begin{equation}
 G_dV=G_d\begin{pmatrix}
 \bar{x}\ones_M+\Delta\bar X_M\\
 \bar{y}\ones_M+\Delta\bar Y_M
 \end{pmatrix}
 =\begin{pmatrix}
 \bar{y}\ones_M-\Delta\bar X_M\\
 \bar{x}\ones_M-\Delta\bar Y_M
 \end{pmatrix}.
 \label{eq:centralidentity}
\end{equation}
The corresponding local Grover action for an arbitrary partition into blocks is derived in Appendix~\ref{app:blocks}.

This extends the $d=4$ illustration of Sec.~II to arbitrary even degree: the two block means are exchanged, while each zero-sum deviation changes sign. The Grover coin satisfies
\begin{equation}
 \Pi G_d\Pi^\dagger=G_d
 \label{eq:groverperm}
\end{equation}
for every permutation matrix $\Pi$; its local action is therefore independent of the numerical channel labels.

\subsection{Orthogonal direct-sum structure and completeness}

We now separate the mean-plus-deviation representation into four contributions:
\begin{equation}
\begin{aligned}
 V={}&\bar{x}\begin{pmatrix}\ones_M\\\zeros_M\end{pmatrix}
 +\begin{pmatrix}\Delta\bar X_M\\\zeros_M\end{pmatrix}\\
 &+\bar{y}\begin{pmatrix}\zeros_M\\\ones_M\end{pmatrix}
 +\begin{pmatrix}\zeros_M\\\Delta\bar Y_M\end{pmatrix}.
\end{aligned}
\label{eq:foursectors}
\end{equation}
These terms belong to the subspaces
\begin{equation}
\begin{aligned}
 \mathcal P_R&=\operatorname{span}\left\{\begin{pmatrix}\ones_M\\\zeros_M\end{pmatrix}\right\},\\
 \mathcal P_L&=\operatorname{span}\left\{\begin{pmatrix}\zeros_M\\\ones_M\end{pmatrix}\right\},
\end{aligned}
\label{eq:Pspaces}
\end{equation}
\begin{equation}
\begin{aligned}
 \mathcal T_L&=\left\{\begin{pmatrix}w\\\zeros_M\end{pmatrix}:w\in\mathbb C^M,\ \ones_M^\dagger w=0\right\},\\
 \mathcal T_R&=\left\{\begin{pmatrix}\zeros_M\\w\end{pmatrix}:w\in\mathbb C^M,\ \ones_M^\dagger w=0\right\}.
\end{aligned}
\label{eq:Tspaces}
\end{equation}

Within the same block, the uniform mean component is orthogonal to every zero-sum vector. Let $c\in\mathbb C$ denote the uniform amplitude ($c=\bar{x}$ or $\bar{y}$ for the corresponding block of $V$), and let $w$ be any zero-sum vector. Then
\begin{equation}
 (c\ones_M)^\dagger w=c^*\ones_M^\dagger w=0.
 \label{eq:orthuniformzero}
\end{equation}
The two uniform components are orthogonal because they occupy different blocks:
\begin{equation}
 \begin{pmatrix}\bar{x}\ones_M\\\zeros_M\end{pmatrix}^{\!\dagger}
 \begin{pmatrix}\zeros_M\\\bar{y}\ones_M\end{pmatrix}=0.
 \label{eq:orthuniformblocks}
\end{equation}
Likewise, for any zero-sum $w_L$ and $w_R$,
\begin{equation}
 \begin{pmatrix}w_L\\\zeros_M\end{pmatrix}^{\!\dagger}
 \begin{pmatrix}\zeros_M\\w_R\end{pmatrix}=0.
 \label{eq:orthzeroblocks}
\end{equation}
The same disjoint-block argument makes a uniform component in one block orthogonal to a zero-sum component in the other block.

The four subspaces are therefore mutually orthogonal, with dimensions\footnote{Each propagation vector is determined by one complex coefficient multiplying $\ones_M$ in its occupied block. For a trapping block, the condition $\sum_{i=1}^{M}w_i=0$ fixes $w_M=-\sum_{i=1}^{M-1}w_i$, leaving $M-1$ independent complex coefficients. The corresponding dimensions over $\mathbb C$ are therefore $1$ and $M-1$, respectively.}
\begin{equation}
\begin{aligned}
 \dim\mathcal P_R=\dim\mathcal P_L&=1,\\
 \dim\mathcal T_L=\dim\mathcal T_R&=M-1.
\end{aligned}
\label{eq:dimensions}
\end{equation}
Their dimensions sum to $2M=d$, giving the complete decomposition
\begin{equation}
 \mathbb C^d=\mathcal P_R\oplus\mathcal T_L\oplus\mathcal P_L\oplus\mathcal T_R.
 \label{eq:directsum}
\end{equation}
Every local vector consequently has a unique decomposition into these four contributions. For $d=4$, all four subspaces are one-dimensional; for $d=6$, each trapping subspace has dimension two. Under the local Grover coin, the two uniform subspaces are exchanged, whereas each zero-sum subspace acquires a minus sign. Their propagation and trapping interpretation follows from combining this local action with the specified shift.

\subsection{Superposition, spatial overlap, and sector identity}

Under the routing shown in Fig.~\ref{fig:evolution}, the shift carries each outgoing block to the corresponding channels at the neighboring vertex. At the next coin operation, a uniform input is transferred to the complementary block and continues along the path. A zero-sum input retains its block with reversed sign, after which the shift returns it to the preceding vertex. Repetition produces ballistic propagation of the mean contributions and two-site motion of the deviations.

Let the initial state be localized at the origin and decomposed as
\begin{equation}
 \Psi(0)=\Psi_{P_R}(0)+\Psi_{T_L}(0)+\Psi_{P_L}(0)+\Psi_{T_R}(0).
 \label{eq:superposition}
\end{equation}
Linearity gives $\Psi(t)=\sum_\mu U^t\Psi_\mu(0)$, with $\mu\in\{P_R,T_L,P_L,T_R\}$. For this routing, the contributions remain locally orthogonal whenever they occupy the same vertex. Contributions from different sectors have zero inner product when they occupy different blocks. Within the same block, a uniform mean component is orthogonal to a zero-sum deviation. Consequently,
\begin{equation}
 \|\Psi_n(t)\|^2=\sum_\mu\|\Psi_{\mu,n}(t)\|^2,
 \label{eq:probaddlocal}
\end{equation}
where $\Psi_{\mu,n}(t)$ denotes the local coin vector of contribution $\mu$ at vertex $n$. This probability additivity underlies the four partial weights introduced in Sec.~\ref{sec:probabilities}.

\section{Four partial probabilities and inverse state design}
\label{sec:probabilities}
\label{sec:inverse}

\subsection{Partial probabilities for $d=4$}

Let
\begin{equation}
 C=\sum_{i=1}^{4}|z_i|^2,
 \label{eq:C4}
\end{equation}
with $C>0$, and define $\bar{x}=(z_1+z_2)/2$ and $\bar{y}=(z_3+z_4)/2$. With the adopted convention, the first pair contributes right propagation and left trapping, whereas the second contributes left propagation and right trapping. Thus
\begin{equation}
 P_R=\frac{2}{C}|\bar{x}|^2,\qquad
 T_L=\frac{1}{C}\left(|z_1-\bar{x}|^2+|z_2-\bar{x}|^2\right),
 \label{eq:PRTL4}
\end{equation}
\begin{equation}
 P_L=\frac{2}{C}|\bar{y}|^2,\qquad
 T_R=\frac{1}{C}\left(|z_3-\bar{y}|^2+|z_4-\bar{y}|^2\right).
 \label{eq:PLTR4}
\end{equation}

For the two blocks, we obtain
\begin{equation}
\begin{aligned}
 P_R+T_L&=\frac{|z_1|^2+|z_2|^2}{C}=\frac{\|X\|^2}{C},\\[3pt]
 P_L+T_R&=\frac{|z_3|^2+|z_4|^2}{C}=\frac{\|Y\|^2}{C}.
\end{aligned}
\label{eq:blocknorms4}
\end{equation}
Adding the two identities explicitly verifies normalization:
\begin{equation}
 P_R+T_L+P_L+T_R=\frac{\sum_{i=1}^4|z_i|^2}{C}=1.
 \label{eq:normalization4}
\end{equation}

\subsection{Generalization to arbitrary even $d=2M$}

For
\begin{equation}
 V=\bigl((z_1,\ldots,z_M),(z_{M+1},\ldots,z_{2M})\bigr)^T,
 \label{eq:generalstate2M}
\end{equation}
define
\begin{equation}
 \bar{x}=\frac1M\sum_{i=1}^{M}z_i,\qquad
 \bar{y}=\frac1M\sum_{i=M+1}^{2M}z_i,\qquad
 C=\sum_{i=1}^{2M}|z_i|^2.
 \label{eq:generalmeansC}
\end{equation}
Then
\begin{equation}
 P_R=\frac{M|\bar{x}|^2}{C},\qquad
 P_L=\frac{M|\bar{y}|^2}{C},
 \label{eq:PRPLgeneral}
\end{equation}
\begin{equation}
 T_L=\frac1C\sum_{i=1}^{M}|z_i-\bar{x}|^2,\qquad
 T_R=\frac1C\sum_{i=M+1}^{2M}|z_i-\bar{y}|^2.
 \label{eq:TLTRgeneral}
\end{equation}
From the preceding expressions, we obtain
\begin{equation}
 \frac{\|X\|^2}{C}=P_R+T_L,\qquad
 \frac{\|Y\|^2}{C}=P_L+T_R,
 \label{eq:blocknorms}
\end{equation}
and therefore
\begin{equation}
 \boxed{P_R+T_L+P_L+T_R=1}.
 \label{eq:normalization}
\end{equation}

\subsection{Inverse design: the general solution}

We now solve Eqs.~\eqref{eq:PRPLgeneral} and~\eqref{eq:TLTRgeneral} for the complex amplitudes $z_i$, taking the four probabilities as prescribed real parameters. We group them according to the two blocks:
\[
\bigl(\underbrace{(P_R,T_L)}_{\text{first block}},\underbrace{(P_L,T_R)}_{\text{second block}}\bigr).
\]
For a normalized initial state, $P_R$ and $T_L$ are determined by the amplitudes in the first block, whereas $P_L$ and $T_R$ are determined by those in the second block. The prescribed probabilities satisfy
\begin{equation}
\begin{gathered}
 P_R,T_L,P_L,T_R\in[0,1],\\
 P_R+T_L+P_L+T_R=1.
\end{gathered}
\label{eq:targetconditions}
\end{equation}
For a normalized state, their general solution is
\begin{equation}
\begin{aligned}
 z_i&=\sqrt{\frac{P_R}{M}}e^{j\phi_R}+\sqrt{T_L}\,\widehat\tau_{L,i},
 &&1\le i\le M,\\[6pt]
 z_i&=\sqrt{\frac{P_L}{M}}e^{j\phi_L}+\sqrt{T_R}\,\widehat\tau_{R,i-M},
 &&M+1\le i\le2M,
\end{aligned}
\label{eq:inversecomponents}
\end{equation}
where $j^2=-1$. The first term in each line is the common block mean. Its magnitude is fixed by the propagation probability; its phase $\phi_R$ or $\phi_L$ is arbitrary. The second term is a component of a zero-sum deviation vector whose magnitude (norm) is fixed by the corresponding trapping probability. The vectors $\widehat\tau_L,\widehat\tau_R\in\mathbb C^M$ can be chosen independently, subject to
\begin{equation}
\begin{gathered}
 \ones_M^\dagger\widehat\tau_L=\ones_M^\dagger\widehat\tau_R=0,\\[2pt]
 \|\widehat\tau_L\|=\|\widehat\tau_R\|=1.
\end{gathered}
\label{eq:taus}
\end{equation}
Collecting these amplitudes into the two blocks gives
\begin{equation}
 V=\begin{pmatrix}
 \sqrt{P_R/M}\,e^{j\phi_R}\ones_M+\sqrt{T_L}\,\widehat\tau_L\\[5pt]
 \sqrt{P_L/M}\,e^{j\phi_L}\ones_M+\sqrt{T_R}\,\widehat\tau_R
 \end{pmatrix}.
 \label{eq:inversegeneral}
\end{equation}
For an unnormalized vector with squared norm $C$, multiply the right-hand sides of Eqs.~\eqref{eq:inversecomponents} and~\eqref{eq:inversegeneral} by $\sqrt C$. Equation~\eqref{eq:inversegeneral} is the general inverse solution. The prescribed probabilities determine the squared norms of the four mutually orthogonal components. The propagation phases are arbitrary, and the unit zero-sum vectors can be chosen freely subject to Eq.~\eqref{eq:taus}.

\subsection{A convenient symmetric representative}
\label{sec:symmetric}

The general solution allows any unit zero-sum deviation pattern. A simple explicit choice places the deviations at equal distances from the block mean, with equally spaced angles in the complex plane. The mean terms remain those of Eq.~\eqref{eq:inversegeneral}; only the unit zero-sum deviation vectors are specified.

Let
\[
\omega_M=e^{j\,2\pi/M},
\]
and define
\begin{equation}
 \rv_M=\left(\omega_M^0,\omega_M^1,\omega_M^2,\ldots,\omega_M^{M-1}\right)^T.
 \label{eq:omega}
\end{equation}
It satisfies $\ones_M^\dagger\rv_M=0$ and $\|\rv_M\|=\sqrt M$. The symmetric family is obtained from the general solution in Eq.~\eqref{eq:inversegeneral} by taking the unit zero-sum deviation vectors to be
\begin{equation}
 \widehat\tau_L=\frac{e^{j\beta}}{\sqrt M}\rv_M,\qquad
 \widehat\tau_R=\frac{e^{j\delta}}{\sqrt M}\rv_M.
 \label{eq:symdirections}
\end{equation}
The factors $e^{j\beta}$ and $e^{j\delta}$ rotate the deviation patterns as a whole. The component formulas become
\begin{equation}
\begin{aligned}
 z_i&=\sqrt{\frac{P_R}{M}}e^{j\phi_R}
 +\sqrt{\frac{T_L}{M}}e^{j[\beta+2\pi(i-1)/M]},\\
 &\hspace{35mm}1\le i\le M,
\end{aligned}
\label{eq:symfirst}
\end{equation}
\begin{equation}
\begin{aligned}
 z_i&=\sqrt{\frac{P_L}{M}}e^{j\phi_L}
 +\sqrt{\frac{T_R}{M}}e^{j[\delta+2\pi(i-M-1)/M]},\\
 &\hspace{35mm}M+1\le i\le2M.
\end{aligned}
\label{eq:symsecond}
\end{equation}
Equivalently, the mean-plus-deviation block form is
\begin{equation}
 V_{\mathrm{sym}}=\frac1{\sqrt M}
 \begin{pmatrix}
 \sqrt{P_R}\,e^{j\phi_R}\ones_M+\sqrt{T_L}\,e^{j\beta}\rv_M\\[5pt]
 \sqrt{P_L}\,e^{j\phi_L}\ones_M+\sqrt{T_R}\,e^{j\delta}\rv_M
 \end{pmatrix}.
 \label{eq:inversesymmetric}
\end{equation}

Separating the mean and zero-sum deviation terms gives the four sector contributions:
\begin{equation}
\begin{aligned}
V_{\mathrm{sym}}={}&
\underbrace{\sqrt{\frac{P_R}{M}}e^{j\phi_R}
\begin{pmatrix}\ones_M\\\zeros_M\end{pmatrix}}_{V_{P_R}}
+\underbrace{\sqrt{\frac{T_L}{M}}e^{j\beta}
\begin{pmatrix}\rv_M\\\zeros_M\end{pmatrix}}_{V_{T_L}}\\[4pt]
&+\underbrace{\sqrt{\frac{P_L}{M}}e^{j\phi_L}
\begin{pmatrix}\zeros_M\\\ones_M\end{pmatrix}}_{V_{P_L}}
+\underbrace{\sqrt{\frac{T_R}{M}}e^{j\delta}
\begin{pmatrix}\zeros_M\\\rv_M\end{pmatrix}}_{V_{T_R}}.
\end{aligned}
\label{eq:symmetricsectors}
\end{equation}
These are the weighted initial components of the four sectors. This representation will be used to prepare the numerical study.

All four phases $\phi_R,\phi_L,\beta,\delta$ are arbitrary. For $M=2$, the only freedom in a unit zero-sum deviation vector is a common phase:
\[
\begin{aligned}
\frac{e^{j\theta}}{\sqrt2}\rv_2
&=\frac{e^{j\theta}}{\sqrt2}(\omega_2^0,\omega_2^1)^T\\[2pt]
&=\frac{e^{j\theta}}{\sqrt2}(1,-1)^T,\qquad\theta\in\mathbb R.
\end{aligned}
\]
The symmetric family therefore covers the full general solution in this case. For $M>2$, the zero-sum and unit-norm conditions allow additional deviation patterns, so the symmetric family represents only a subset of the general solution.

\section{Explicit inverse-design examples}
\label{sec:examples}

Table~\ref{tab:examples} gives representative normalized initial states for $d=4$ and $d=6$, obtained from the symmetric solution with $\phi_R=\phi_L=\beta=\delta=0$. The target probabilities are grouped as $\bigl((P_R,T_L),(P_L,T_R)\bigr)$, with one pair for each block.

For $M=2$, the vectors used in this representation are
\[
\begin{gathered}
\ones_2=(1,1)^T,\qquad \zeros_2=(0,0)^T,\\[2pt]
\rv_2=(\omega_2^0,\omega_2^1)^T=(1,-1)^T,\qquad\omega_2=-1.
\end{gathered}
\]
For $M=3$,
\[
\begin{gathered}
\ones_3=(1,1,1)^T,\qquad \zeros_3=(0,0,0)^T,\\[2pt]
\rv_3=(\omega_3^0,\omega_3^1,\omega_3^2)^T=(1,\omega_3,\omega_3^2)^T,
\end{gathered}
\]
where
\[
\omega_3=e^{j\,2\pi/3}=-\frac12+j\frac{\sqrt3}{2}.
\]

\begin{table*}[!t]
\caption{Representative normalized initial states realizing the target probabilities in the first column. Inner parentheses identify the two blocks. All four phases are zero. The last $d=4$ state is used in Sec.~\ref{sec:numerical}.}
\label{tab:examples}
\centering
\fontsize{8.3}{10}\selectfont
\setlength{\tabcolsep}{2pt}
\renewcommand{\arraystretch}{1.1}
\begin{tabular*}{\textwidth}{@{\extracolsep{\fill}}ccc@{}}
\hline\hline
\shortstack{Target probabilities\\$\bigl((P_R,T_L),(P_L,T_R)\bigr)$}
& $d=4$ ($M=2$): normalized $V$ & $d=6$ ($M=3$): normalized $V$ \\[3pt]
\hline\noalign{\vskip4pt}
$\bigl((1,0),(0,0)\bigr)$
& $\dfrac1{\sqrt2}\tabblocks{1,1}{0,0}$
& $\dfrac1{\sqrt3}\tabblocks{1,1,1}{0,0,0}$\\[6pt]
$\bigl((0,1),(0,0)\bigr)$
& $\dfrac1{\sqrt2}\tabblocks{1,-1}{0,0}$
& $\dfrac1{\sqrt3}\tabblocks{1,\omega_3,\omega_3^2}{0,0,0}$\\[6pt]
$\bigl((0,0),(1,0)\bigr)$
& $\dfrac1{\sqrt2}\tabblocks{0,0}{1,1}$
& $\dfrac1{\sqrt3}\tabblocks{0,0,0}{1,1,1}$\\[6pt]
$\bigl((0,0),(0,1)\bigr)$
& $\dfrac1{\sqrt2}\tabblocks{0,0}{1,-1}$
& $\dfrac1{\sqrt3}\tabblocks{0,0,0}{1,\omega_3,\omega_3^2}$\\[6pt]
$\bigl((\tfrac14,\tfrac14),(\tfrac14,\tfrac14)\bigr)$
& $\dfrac1{\sqrt2}\tabblocks{1,0}{1,0}$
& $\dfrac1{\sqrt{12}}\tabblocks{2,1+\omega_3,1+\omega_3^2}{2,1+\omega_3,1+\omega_3^2}$\\[6pt]
$\bigl((0.4,0.1),(0.3,0.2)\bigr)$
& $\dfrac1{\sqrt{20}}\tabblocks{3,1}{\sqrt3+\sqrt2,\sqrt3-\sqrt2}$
& $\dfrac1{\sqrt{30}}\tabblocks{3,2+\omega_3,2+\omega_3^2}{\sqrt3+\sqrt2,\sqrt3+\sqrt2\omega_3,\sqrt3+\sqrt2\omega_3^2}$\\[7pt]
\hline\hline
\end{tabular*}
\end{table*}

\section{Numerical verification of the analytical four-sector dynamics}
\label{sec:numerical}

The preceding analytical results establish the orthogonality and completeness of the four sectors. The numerical study illustrates their evolution and checks the predicted probabilities, including when different sector contributions reach the same vertex. We evolve the total initial state and its four weighted sector components separately under the same walk operator. At each vertex and time step, we compare the total probability with the sum of the four sector probabilities through $\Delta(n,t)$, defined in Eq.~\eqref{eq:residual}. This difference is analytically zero. In the simulation, its largest absolute value is of order $10^{-16}$, consistent with the finite precision of the numerical calculation.

\subsection{Initial-state decomposition into the four sectors}

For the numerical study, we take $d=4$, corresponding to $M=2$, set all four phases to zero, and prescribe
\[
\bigl((P_R,T_L),(P_L,T_R)\bigr)
=\bigl((0.4,0.1),(0.3,0.2)\bigr).
\]
Substituting these values into the four-sector representation in Eq.~\eqref{eq:symmetricsectors} gives the weighted initial components
\begin{equation}
\begin{aligned}
\vt={}&
\underbrace{\frac2{\sqrt{20}}\stackblocks{\pair{1}{1}}{\pair{0}{0}}}_{V_{P_R}}
+\underbrace{\frac1{\sqrt{20}}\stackblocks{\pair{1}{-1}}{\pair{0}{0}}}_{V_{T_L}}\\[4pt]
&+\underbrace{\frac{\sqrt3}{\sqrt{20}}\stackblocks{\pair{0}{0}}{\pair{1}{1}}}_{V_{P_L}}
+\underbrace{\frac{\sqrt2}{\sqrt{20}}\stackblocks{\pair{0}{0}}{\pair{1}{-1}}}_{V_{T_R}}.
\end{aligned}
\label{eq:exampleweighted4}
\end{equation}
Their squared norms reproduce the prescribed probabilities:
\begin{equation}
\begin{aligned}
\|V_{P_R}\|^2&=\frac4{20}(1+1)=0.4,\\[2pt]
\|V_{T_L}\|^2&=\frac1{20}(1+1)=0.1,\\[2pt]
\|V_{P_L}\|^2&=\frac3{20}(1+1)=0.3,\\[2pt]
\|V_{T_R}\|^2&=\frac2{20}(1+1)=0.2.
\end{aligned}
\label{eq:exampleweights4}
\end{equation}
Since the four components are mutually orthogonal,
\begin{equation}
\begin{aligned}
\|\vt\|^2
&=\|V_{P_R}\|^2+\|V_{T_L}\|^2\\
&\quad+\|V_{P_L}\|^2+\|V_{T_R}\|^2=1.
\end{aligned}
\label{eq:examplenorm4}
\end{equation}
We place the initial state at the origin and define
\begin{equation}
\begin{aligned}
|\Psi_{\mathrm{tot}}(0)\rangle&=|0\rangle\otimes\vt,\\[2pt]
|\Psi_\mu(0)\rangle&=|0\rangle\otimes V_\mu,
\end{aligned}
\label{eq:exampleinitialstates}
\end{equation}
where $\mu\in\{P_R,T_L,P_L,T_R\}$. Thus,
\begin{equation}
\begin{aligned}
|\Psi_{\mathrm{tot}}(0)\rangle={}&|\Psi_{P_R}(0)\rangle+|\Psi_{T_L}(0)\rangle\\
&+|\Psi_{P_L}(0)\rangle+|\Psi_{T_R}(0)\rangle.
\end{aligned}
\label{eq:exampleketsplit}
\end{equation}
Linearity gives
\begin{equation}
\begin{aligned}
|\Psi_\mu(t)\rangle&=U^t|\Psi_\mu(0)\rangle,\\[3pt]
|\Psi_{\mathrm{tot}}(t)\rangle&=U^t|\Psi_{\mathrm{tot}}(0)\rangle
=\sum_\mu|\Psi_\mu(t)\rangle.
\end{aligned}
\label{eq:exampleevolution}
\end{equation}
All five evolutions therefore use the same walk operator with different initial states. Each sector component retains the initial weight specified in Eq.~\eqref{eq:exampleweights4}.

\subsection{Simulation protocol and diagnostics}

We simulate the $d=4$ walk on $N=9$ vertices for $T=20$ steps. The vertices are labeled $n=-4,\ldots,4$. Periodic boundary conditions connect vertices $-4$ and $+4$. This periodic closure forms a multiedge cycle. The local coin is $G_4$, and the shift follows the routing defined in Sec.~II. One time step is described by
\begin{equation}
U=S\bigl(I_9\otimes G_4\bigr).
\label{eq:numericalU}
\end{equation}
The local initial state in Eq.~\eqref{eq:exampleweighted4} has the numerical form
\begin{equation}
V_{\mathrm{num}}=\vt\approx
\stackblocks{\pair{0.670820}{0.223607}}{\pair{0.703526}{0.071071}}.
\label{eq:numstate}
\end{equation}
The simulation uses the amplitudes at the full precision of the numerical calculation (approximately 16 significant digits). The values displayed in Eq.~\eqref{eq:numstate} are rounded to six decimal places for readability.

At each vertex $n$ and time step $t$, we calculate
\begin{equation}
\begin{aligned}
p_\mu(n,t)&=\left\|\Pi_n|\Psi_\mu(t)\rangle\right\|^2,\\[2pt]
p_{\mathrm{tot}}(n,t)&=\left\|\Pi_n|\Psi_{\mathrm{tot}}(t)\rangle\right\|^2,
\end{aligned}
\label{eq:sectorprob}
\end{equation}
where $\Pi_n=|n\rangle\langle n|\otimes I_4$. These probabilities include the contributions of all four coin channels at the selected vertex. Summing over all vertices gives the total weight of each sector:
\begin{equation}
W_\mu(t)=\sum_n p_\mu(n,t).
\label{eq:sectorweight}
\end{equation}
To compare the total probability with the sum of the four sector probabilities, we calculate their difference at each vertex and time step:
\begin{equation}
\Delta(n,t)=p_{\mathrm{tot}}(n,t)-\sum_\mu p_\mu(n,t).
\label{eq:residual}
\end{equation}
The step operator satisfies $U^\dagger U=I$, so the norm of each state is preserved during evolution. Throughout all 20 steps, the calculated sector weights remain at $0.40$, $0.10$, $0.30$, and $0.20$, respectively, to the precision of the calculation. The largest absolute probability difference over all vertices and time steps is
\begin{equation}
\max_{n,t}|\Delta(n,t)|\simeq3.33\times10^{-16}.
\label{eq:maxresidual}
\end{equation}
This difference is consistent with the small rounding errors introduced when calculations use a finite number of digits. Its numerical value may vary slightly with the simulation
parameters, whereas the analytical difference is exactly zero. The total probability map therefore agrees with the sum of the four sector maps to the precision of the calculation.

\subsection{Sector-resolved dynamics and spatial overlap}

\begin{figure}[!t]
\centering
\includegraphics[width=\columnwidth]{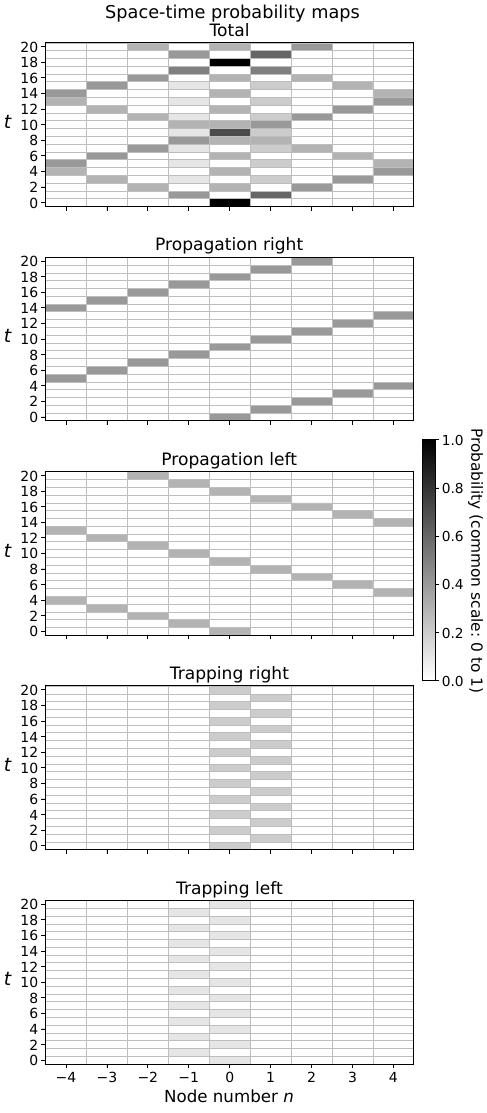}
\caption{Total and sector-resolved space-time probabilities for $d=4$, $N=9$, and $T=20$ with periodic routing. The designed weights are $\bigl((P_R,T_L),(P_L,T_R)\bigr)=\bigl((0.40,0.10),(0.30,0.20)\bigr)$. All panels share a grayscale from white ($0$) to black ($1$). The first complete spatial reunion occurs at $t=18$. The maximum absolute difference between the total probability and the sum of the four sector probabilities is approximately $3.33\times10^{-16}$.}
\label{fig:numerical}
\end{figure}

Figure~\ref{fig:numerical} shows the total probability and the four sector probabilities. All five panels use the same grayscale, from white for zero probability to black for unit probability, so their intensities can be compared directly. The propagating sectors move one vertex per step in opposite directions, while each trapping sector alternates between the origin and one adjacent vertex.

For the adopted routing and the vertex labels $n=-4,\ldots,4$, the analytical positions are
\begin{equation}
\begin{aligned}
n_{P_R}(t)&=\bigl((t+4)\bmod9\bigr)-4,\\[2pt]
n_{P_L}(t)&=\bigl((-t+4)\bmod9\bigr)-4,\\[2pt]
n_{T_L}(t)&=-\frac{1-(-1)^t}{2},\\[2pt]
n_{T_R}(t)&=\frac{1-(-1)^t}{2}.
\end{aligned}
\label{eq:periodicpositions}
\end{equation}
The first two expressions include the passage between $+4$ and $-4$ at the periodic boundary. The trapping sectors alternate between $0$ and $-1$ for $T_L$, and between $0$ and $+1$ for $T_R$.

At $t=1$, the $P_R$ and $T_R$ contributions occupy $n=+1$, giving a total probability of $0.60$. The $P_L$ and $T_L$ contributions occupy $n=-1$, giving a total probability of $0.40$.

For this routing, sector contributions that meet at the same vertex remain locally orthogonal. Contributions from different sectors have zero inner product when they occupy different blocks. Within the same block, a uniform mean component is orthogonal to a zero-sum deviation. Consequently, the vertex probability is
\begin{equation}
\begin{aligned}
p_{\mathrm{tot}}(n,t)={}&p_{P_R}(n,t)+p_{T_L}(n,t)\\
&+p_{P_L}(n,t)+p_{T_R}(n,t).
\end{aligned}
\label{eq:numericaladditivity}
\end{equation}
The numerical comparison in Eq.~\eqref{eq:maxresidual} confirms this analytical identity to the precision of the calculation. Repeated encounters on the periodic graph preserve the trajectories and weights of the four sectors.

For a periodic graph with $N$ vertices, both propagating sectors return to the origin at times that are multiples of $N$. Both trapping sectors are at the origin at even times. With all four sector weights nonzero and $N\ge3$, the first complete spatial reunion at the origin occurs at the smallest positive time that is both a multiple of $N$ and an even integer:
\begin{equation}
t_{\mathrm{reunion}}=\operatorname{lcm}(N,2)
=\begin{cases}
N,&N\ \text{even},\\
2N,&N\ \text{odd}.
\end{cases}
\label{eq:reunion}
\end{equation}
For $N=9$, the propagating sectors return to $n=0$ at $t=9$, while the trapping sectors are then at $n=-1$ and $n=+1$. At $t=18$, all four sectors return to the origin, where $p_{\mathrm{tot}}(0,18)=1$. This is the first complete spatial reunion in the simulation.

\section{Conclusion}
\label{sec:conclusion}

We have established a unified four-sector description of directional propagation and bounded trapping within a single Grover-coined quantum walk on a multiedge path graph. Every local initial state admits a unique orthogonal decomposition into two uniform mean components and two zero-sum deviation components. The compatibility between the graph's local partition into two equal directional blocks and the Grover coin produces a pure exchange of the block means and a sign reversal of the deviations. In this walk on the multiedge path graph, the mean components propagate in opposite directions, while the zero-sum deviations remain trapped on either side of the launch vertex.

The inverse construction provides a family of initial states realizing any prescribed set of nonnegative sector probabilities that sum to unity, while keeping the graph and the walk operator fixed. Previous quantum-walk studies have explored transport control through changes to coin operations or graph connectivity~\cite{Zhan2014,Nitsche2016}. For the routing studied here, the four sector contributions retain their weights and remain locally orthogonal even when they occupy the same vertex. The numerical study reproduces the predicted motion and local probability additivity in the periodically closed implementation, including repeated spatial overlaps.

Appendix~\ref{app:blocks} extends the local mean--deviation description to arbitrary finite degree $d$ and partitions into any number of blocks, whose dimensions $p_k$ may differ and satisfy $\sum_k p_k=d$. The mean of block $k$ contributes to every other block mean with coefficient $2p_k/d$, while its contribution retained in the original block has coefficient $-(1-2p_k/d)$. Each zero-sum deviation remains within its original block and has its sign reversed. The retained contribution of a block mean vanishes when $p_k=d/2$. Two equal blocks therefore give pure mean exchange.

Revisiting previously studied quantum walks through the mean--deviation decomposition may reveal features less apparent in their original descriptions and clarify the contributions of the mean and deviation components to the observed dynamics. For future studies on other graph topologies, the general local relation provides a starting point for identifying graph routings compatible with the desired transport and confinement properties. The local action of the coin and its compatibility with the shift must be considered together to determine how these components evolve across the graph.

\begin{acknowledgments}
The author used OpenAI ChatGPT (GPT-6.1 Sol and Astra), under his direction, to assist with manuscript preparation, literature search and organization, and the development and checking of plotting code for his numerical simulations. The simulations were based on the author's original code. OpenAI ChatGPT assisted with the plotting routines. The author reviewed the outputs, verified the references and numerical results, and takes full responsibility for the manuscript.
\end{acknowledgments}

\begin{samepage}
\section*{Data availability}
No external datasets were used. The model definitions, initial states, and numerical parameters needed to reproduce the reported results are given in this article.
The Python simulation code and numerical data underlying Fig.~\ref{fig:numerical} are openly available in Zenodo~\cite{Bougroura2026Code}.
\end{samepage}

\appendix
\section{Local Grover action for an arbitrary partition into blocks}
\label{app:blocks}

Let a local vector $V\in\mathbb C^d$ be partitioned into $n$ nonempty blocks $X_k\in\mathbb C^{p_k}$, where $p_k$ is the dimension of block $k$ and
\[
V=\begin{pmatrix}X_1\\X_2\\\vdots\\X_n\end{pmatrix},
\qquad \sum_{k=1}^{n}p_k=d.
\]
The dimensions may be equal or unequal. For each block, define its mean and zero-sum deviation by
\[
\begin{aligned}
\bar x_k&=\frac{1}{p_k}\ones_{p_k}^{\dagger}X_k,\\[2pt]
\Delta\bar X_{p_k}^{(k)}&=X_k-\bar x_k\ones_{p_k},\\[2pt]
\ones_{p_k}^{\dagger}\Delta\bar X_{p_k}^{(k)}&=0.
\end{aligned}
\]
The index $k$ identifies the block, and $p_k$ specifies its dimension. The superscript $(k)$ distinguishes deviation vectors belonging to different blocks, including blocks of the same dimension. Thus,
\[
X_k=\bar x_k\ones_{p_k}+\Delta\bar X_{p_k}^{(k)}.
\]
Because the deviations have zero sum, the total amplitude sum is
\[
s=\sum_{i=1}^{n}p_i\bar x_i.
\]
Applying the componentwise Grover rule $z'_i=2s/d-z_i$ to block $k$ gives
\[
\begin{aligned}
X'_k&=\frac{2s}{d}\ones_{p_k}-X_k\\[2pt]
&=\frac{2}{d}\left(\sum_{i=1}^{n}p_i\bar x_i\right)\ones_{p_k}-X_k.
\end{aligned}
\]
Substituting $X_k=\bar x_k\ones_{p_k}+\Delta\bar X_{p_k}^{(k)}$ and separating the contribution of the original block from those of the other blocks yields
\begin{equation}
X'_k=\left[-\left(1-\frac{2p_k}{d}\right)\bar x_k
+\sum_{\substack{i=1\\i\ne k}}^{n}\frac{2p_i}{d}\bar x_i\right]\ones_{p_k}
-\Delta\bar X_{p_k}^{(k)}.
\label{eq:arbitraryblocks}
\end{equation}
The restricted sum includes all block indices from $1$ to $n$ except $k$.

Equation~\eqref{eq:arbitraryblocks} separates two behaviors. The zero-sum deviation of each block remains within that block and changes sign. The mean of source block $k$ contributes $(2p_k/d)\bar x_k$ to the mean of every other block; its contribution in its original block is $-(1-2p_k/d)\bar x_k$.

The retained mean coefficient is negative for $p_k<d/2$, zero for $p_k=d/2$, and positive for $p_k>d/2$. When $p_k=d/2$, a uniform input supported in that block is transferred completely to its complement. For two equal blocks, $p_1=p_2=M=d/2$, Eq.~\eqref{eq:arbitraryblocks} reduces to Eq.~\eqref{eq:centralidentity}: the two means are exchanged and the deviations are sign-reversed. The notation agrees with the main text through
\[
\begin{gathered}
X_1=X,\quad X_2=Y,\quad \bar x_1=\bar x,\quad\bar x_2=\bar y,\\[2pt]
\Delta\bar X_M^{(1)}=\Delta\bar X_M,\qquad
\Delta\bar X_M^{(2)}=\Delta\bar Y_M.
\end{gathered}
\]

The zero-sum deviation space in a block of dimension $p_k$ has dimension $p_k-1$. A nonzero zero-sum deviation is therefore possible precisely when $p_k\ge2$. Spatial propagation and confinement additionally depend on the shift and graph routing.

As a consistency check, take $n=d$ and $p_k=1$ for every block. Each block then contains one component, so all deviations vanish and its mean equals that component. Define the vector of block means by
\[
\bar{\boldsymbol x}=(\bar x_1,\ldots,\bar x_d)^T.
\]
Equation~\eqref{eq:arbitraryblocks} becomes the standard Grover transformation,
\begin{equation}
\begin{aligned}
\bar{\boldsymbol x}'&=
\begin{pmatrix}
-(1-2/d)&2/d&\cdots&2/d\\
2/d&-(1-2/d)&\cdots&2/d\\
\vdots&\vdots&\ddots&\vdots\\
2/d&2/d&\cdots&-(1-2/d)
\end{pmatrix}\bar{\boldsymbol x}
\\[2pt]&=G_d\bar{\boldsymbol x}.
\end{aligned}
\label{eq:singletongrover}
\end{equation}
For three singleton blocks, this gives
\[
G_3=\frac13\begin{pmatrix}-1&2&2\\2&-1&2\\2&2&-1\end{pmatrix}.
\]
Thus the block formula recovers the usual componentwise Grover matrix when the partition consists of singleton blocks.

\end{document}